\documentclass[11pt,a4paper,twoside]{article}
\usepackage{epsfig}
\usepackage{amstex}
\usepackage{amssymb}

\newcommand{\rf}[1]{(\ref{#1})}
\newcommand{\scaption}[1] {\caption{\footnotesize{#1}}}

\begin{document}
\title{The dynamics of a low-order coupled\\ ocean-atmosphere model}
\author{Lennaert van Veen\footnotemark[1]  \and Theo Opsteegh\footnotemark[2] \and Ferdinand Verhulst\footnotemark[1]}
\renewcommand{\thefootnote}{\fnsymbol{footnote}}
\footnotetext[1]{Mathematical Institute, University of Utrecht, PO Box 80.010, 3508 TA Utrecht, the Netherlands}
\footnotetext[2]{Royal Netherlands Meteorological Institute (KNMI), PO Box 201, 3730 AE  de Bilt}

\maketitle

\begin{abstract}
A system of five ordinary differential equations is studied which combines the Lorenz-84 model for the atmosphere and a box model for the ocean. The behaviour of this system is studied as a function of the coupling parameters. For most parameter values, the dynamics of the atmosphere model is dominant. Stable equilibria are found, as well as periodic solutions and chaotic attractors. For a range of parameter values, competing attractors exist. The Kaplan-Yorke dimension and the correlation dimension of the chaotic attractor are numerically calculated and compared to the values found in the uncoupled Lorenz model. The correlation dimension differs much less than te Kaplan-Yorke dimension, indicating that there is little variability in the ocean model. In the transition from periodic behaviour to chaos intermittency is observed. This is explained by means of bifurcation analysis. The intermittent behaviour occurs near a Neimark-Sacker bifurcation at which a periodic solution loses its stability. The average length of a periodic interval in the intermittent regime, $l$, is studied as a function of the bifurcation parameter. Near the bifurcation point it shows a power law scaling. It diverges as $l \propto \epsilon^{-\alpha}$, where $\alpha\approx 0.06$ and $\epsilon$ is the distance from the bifurcation point, in reasonable agreement with the results of Pomeau and Manneville (Commun. Math. Phys. {\bf 74}, 1980). The intermittent behaviour persists beyond the point where the unstable periodic solution disappears in a saddle node bifurcation. The length of the periodic intervals is governed by the time scale of the ocean component. Thus, in this regime the ocean model has a considerable influence on the dynamics of the coupled system.
\end{abstract}

\section{Introduction}

On a time scale of days or weeks, the atmospheric component of the earth's climate system is dominant. Therefore, for short range weather forecasts oceanic variables, such as the sea surface temperature, can be considered fixed. On a much longer time scale, say years or decades, the ocean's dynamics and its coupling to the atmosphere can play an important role. It has to be taken into account when studying for instance decadal climate variability, see e.g. \cite{lati}, or anthropogenic influence like the greenhouse effect. 

For such purposes state-of-the-art climate models are often used, which possess millions of degrees of freedom. Even so-called intermediate models, with a rather coarse resolution by meteorological standards, still have thousands of degrees of freedom. The results of experiments with such models are analysed statistically, as they are out of reach of the ordinary analysis of dynamical systems theory. One important open issue is the interplay of the short time scale of the atmospheric, intrinsically chaotic, components, and the long time scale of the oceanic component. As much understanding of atmosphere models has been gained by looking at extremely low dimensional truncations, our aim is to do the same for coupled models.

A proposal for a low order coupled model is due to Roebber \cite{roeb}. He coupled the Lorenz-84 model, which is a metaphor for the general circulation of the atmosphere \cite{lor2}, to Stommel's box model for a single ocean basin \cite{stom}. Our model is similar to Roebber's, but we have simplified the ocean model somewhat. The feature modeled by Stommel in \cite{stom} is the thermohaline circulation (THC) in the North Atlantic ocean. This is the large scale circulation driven by the north-south heating gradient on one hand, and the difference in salt content of the sea water on the other. The Lorenz model describes the westerly circulation, i.e. the jet stream, and traveling planetary waves.

Experiments with realistic climate models (see, for instance \cite{grot}) indicate that the circulation of the ocean is largely driven by the atmospheric dynamics. In contrast, the feedback to the atmosphere seems to be rather weak, and only notable on long time scales. Therefore, we assume that the coupling terms in the ocean model are of the same order of magnitude as its internal dynamics. The coupling terms in the atmosphere model are taken much smaller than its internal dynamics. The behaviour of the coupled system is then investigated as a function of the coupling parameters in the atmosphere model.

When varying these parameters we find stable equilibrium points, as well as periodic solutions and chaotic attractors. By means of numerical algorithms the Kaplan-Yorke dimension and the correlation dimension of the chaotic attractors are calculated. The difference between the typical Kaplan-Yorke dimension found in the coupled system and the value found in the uncoupled Lorenz system is almost two, the dimension of the ocean model. The difference in correlation dimension is only half as big. This is related to the fact that there is little variability in the ocean model as coupled to the atmosphere model. It is basically a relaxation equation driven by a chaotic forcing.

The main feature of the ocean box model, in fact the reason for studying it in the first place, is the occurrence of coexisting stable equilibria. One of these equilibria describes the temperature driven THC which is currently observed, with warmer water flowing poleward in the upper layer and cooler water flowing back towards the equator in a deeper layer. This circulation is driven by the heating gradient. The other equilibrium describes an inverted THC, driven by the salinity gradient. We show that in the coupled model, for a range of parameter values, there also exists an attracting set in phase space on which the THC is salinity driven. This may be an equilibrium point or a periodic solution. For these parameter values, the model has competing attractors, as there also exists an attracting set on which the THC is temperature driven. This may be a periodic solution or a chaotic attractor.

Another property of the coupled model is the intermittent behaviour, which is observed in the transition from periodic to chaotic motion.
By means of bifurcation analysis of periodic solutions this behaviour can be studied in detail. It turns out, that a periodic solution loses its stability in a Neimark-Sacker bifurcation. Very close to the Neimark-Sacker bifurcation a saddle node bifurcation occurs, at which the periodic solution disappears. The intermittent behaviour persists beyond this point. This phenomenon might be called 'skeleton dynamics' after Nishiura and Ueyama \cite{nish}. Both the Neimark-Sacker and the saddle-node bifurcation are local, which means that some distance away from the bifurcating structure in phase space, the vector field remains essentially the same. The 'ghost' of the periodic orbit keeps attracting the phase point, but only for a finite time. The length of the seemingly periodic intervals can be measured, and we can consider its distribution as a function of the bifurcation parameter. This approach was first taken by Pomeau and Manneville \cite{pome}. They also made a prediction for the order of the divergence of the average length of a periodic interval, $l$, as the bifurcation parameter approaches its critical value. Although analytical arguments suggest a divergence as $\ln 1/\epsilon$, where $\epsilon$ is the distance from the critical value, their own computer simulations showed a power law scaling, i.e. $l \propto \epsilon^{-\alpha}$. The exponent they measure, $\alpha \approx 0.04$, agrees reasonably well with our result, $\alpha \approx 0.06$.

The power law scaling holds only for very small values of $\epsilon$. The intermittent behaviour is found in a much larger range in parameter space. It is our conjecture, that the presence of the slow ocean system enhances the intermittent behaviour. Just beyond the Neimark-Sacker point, during a periodic interval the phase point approaches the unstable periodic solution near its stable manifold. Here, the convergence rate is set by the time scale of the slow system. Thus, the periodic intervals are much longer than the period of the periodic solution, indeed comparable to the relaxation times of the ocean model.

Summarising, we can say that, for a broad range of parameter values, the ocean model does not seem to play a role of importance because of the weak coupling to the atmosphere model.
The atmospheric dynamics is dominant. Two notable exceptions are the occurrence of attracting equilibria and periodic solutions with an inverted THC, also present in the uncoupled ocean model, and the intermittent transition to chaos. The intermittency is generic in the sense that, mathematically, the occurrence of chaos and the loss of stability of periodic orbits through a Neimark-Sacker bifurcation are. Whether it is generic in a hierarchy of increasingly realistic models with increasing dimension remains to be found out.

\section{The Lorenz-84 general circulation model}

Like the Lorenz-63 model, a famous example of a low-order model showing chaotic behaviour, the Lorenz-84 model is a Galerkin truncation of the Navier-Stokes equations. Where the '63 model describes convection, the '84 model gives the simplest approximation to the general atmospheric circulation at midlatitude. The approximation is applicable on a $\beta$-plane\cite{not1}, which we place over the North Atlantic ocean. 

With this derivation in mind, we can give a physical interpretation of the variables of the Lorenz-84 model: $x$ is the intensity of the westerly circulation, $y$ and $z$ are the sine and cosine components of a large traveling wave.
The time derivatives are given by 
\begin{figure}[t]
\begin{picture}(375,220)(0,0)
\epsfysize=200pt
\put(0,0){\epsfig{file=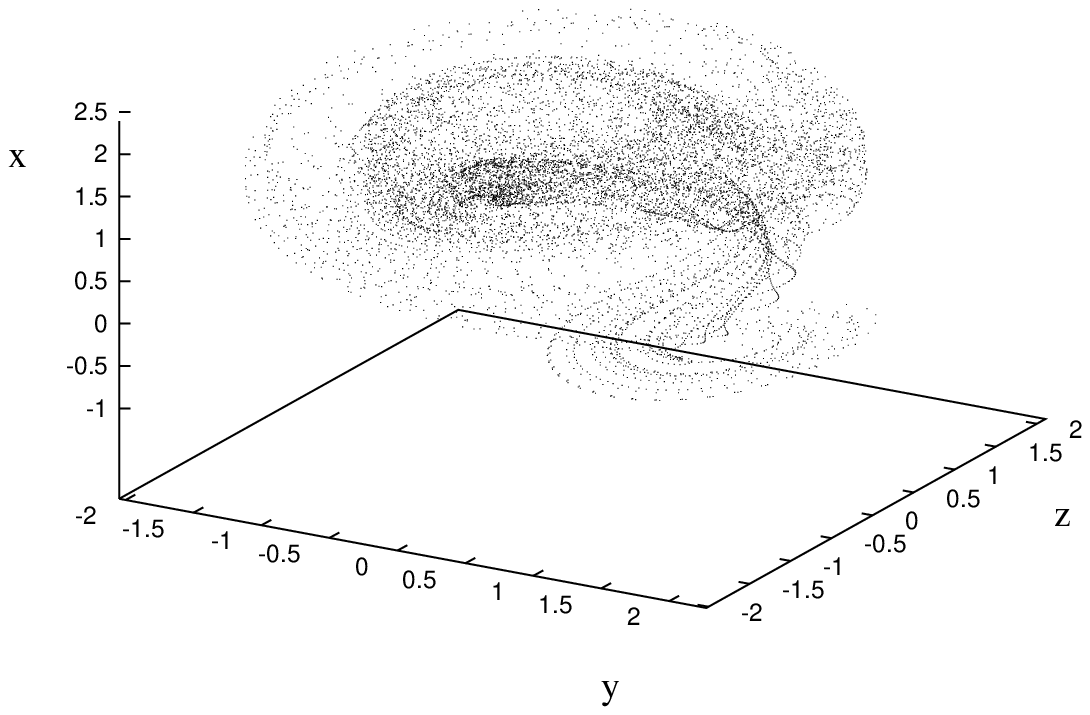}}
\end{picture}
\scaption{Chaotic motion in the Lorenz-84 model for $(F,G)=(8,1)$: about $10^{4}$ points on the attractor.}
\label{lorat}
\end{figure}
\begin{align}
\dot{x} & = -y^{2}-z^{2}-a x + a F 		\label{lor84x} \\
\dot{y} & = x y- b x z - y + G		\label{lor84y} \\
\dot{z} & = b x y + x z - z			\label{lor84z}
\end{align}
where $F$ and $G$ are forcing terms due to the average north-south temperature contrast and the earth-sea temperature contrast, respectively. Conventionally we take $a=1/4$ and $b=4$.

The behaviour of this model has been studied extensively since its introduction by Lorenz \cite{lor2}. Numerical and analytical explorations can be found for instance in \cite{mas2} and \cite{sica}. A bifurcation analysis is presented in \cite{shil}. The bifurcation diagram of this model is quite rich. It brings forth equilibrium points, periodic and quasi periodic orbits as well as chaotic motion. Qualitatively the behaviour can be sketched by looking at the energy transfer between the westerly circulation and the traveling wave. The energy content of the westerly circulation tends to grow, forced by solar heating. Above a certain value however this circulation becomes unstable and energy is transferred to traveling waves, and then dissipated. The energy content of the westerly circulation decreases rapidly and the cycle repeats itself in a periodic or irregular fashion. In figure~\rf{lorat} one can see that the orbit tends to spiral around the $x$-axis towards a critical value of $x$, then drops towards the $y,z$-plane.

At parameter values $(F,G)=(6,1)$ two stable periodic solutions coexist. These parameter values are called summer conditions. For $(F,G)=(8,1)$ the behaviour is chaotic (see figure \rf{lorat}). These parameter values are called winter conditions. If we fix these forcing parameters to summer conditions in the coupled model, described below, no complex dynamics arise. When varying the coupling parameters we see only equilibrium points and periodic solutions. In our investigations we will take $(F,G)=(8,1)$, i.e. we will stick to perpetual winter conditions.

\section{The box model for a single ocean basin}

The ocean-box model was introduced by Stommel in 1961 \cite{stom}. It is a simple model of a single ocean basin, the North Atlantic. This basin is divided in two boxes, one at the equator and one at the north pole. Within the boxes the water is supposed to be perfectly mixed, so that the temperature and salinity are constant within each box but may differ between them. This drives a circulation between the boxes which represents the thermohaline circulation. Water evaporates from the equatorial box and precipitates into the polar box. Thus the salinity difference between the boxes is enhanced. The temperature difference is maintained by the difference in heat flux from the sun. Thus, the salinity and the temperature difference drive a circulation in opposite directions. For a suitable choice of parameters, both the circulation driven by salinity and the circulation driven by temperature occur as stable solutions in this model \cite{stom}. In contrast to the Lorenz model, no complex dynamics arise.

Figure \rf{box} shows the setting of the model. The volume of water is kept equal, but its density may differ between the boxes. Using a linearised equation of state and some assumptions on the damping, dynamical equations for the temperature difference $T=T_{e}-T_{p}$ and the salinity difference $S=S_{e}-S_{p}$ can be derived.   
They are
\begin{align}
 & \dot{T} = k_{a}(T_{a}-T) - |f(T,S)|T - k_{w}T \label{boxeq1} \\
 & \dot{S} = \delta - |f(T,S)|S - k_{w}S \label{boxeq2} \\
 & f = \omega T - \xi S \label{flow} 
\end{align}
where $k_{a}$ is the coefficient of heat exchange between ocean and atmosphere, $k_{w}$ is the coefficient of internal diffusion and $\omega$ and $\xi$ derive from the linearised equation of state. The flow, $f$, represents the THC. It is positive when temperature driven and negative when salinity driven. The inhomogeneous forcing by solar heating and atmospheric water transport are given by $T_{a}$ and $\delta$, respectively. When coupling the box model to the Lorenz-84 model, we will use Roebbers estimates for the parameters in \rf{boxeq1}-\rf{flow}. The volume of the deep ocean box, not present in our model, is simply divided between the polar and the equatorial box \cite{zond}.

The absolute value in \rf{boxeq1} and \rf{boxeq2} was put there by Stommel, arguing that the mixing of the water should be independent of the direction of the flow. A more straightforward derivation of the equations of motion of a simple ocean model related to the box model indicates that this is indeed the case, although the term comes out quadratic instead of piecewise linear \cite{maas}. If we take this term to be quadratic in the coupled model, described below, the average values of $T$ and $S$ change significantly, but we find qualitatively the same behaviour.

\begin{figure}[t]
\begin{picture}(375,170)(0,0)
\epsfxsize=300pt
\centerline{\epsfig{file=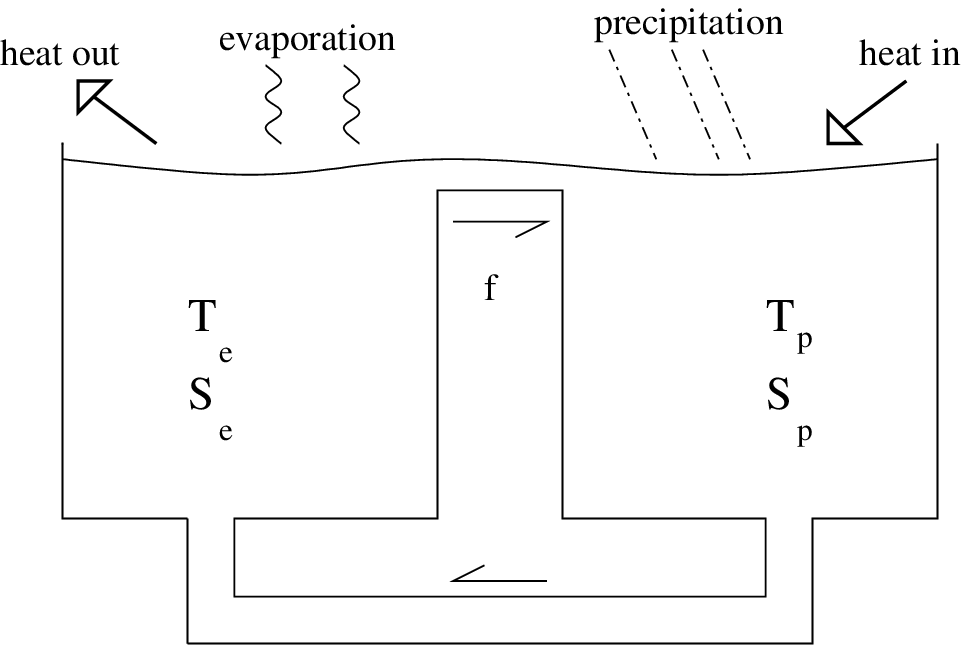}}
\end{picture}
\scaption{The two box model. Water evaporates from the warmer equatorial box on the left and is transported through the atmosphere to the polar box on the right. The flow $f$ is positive when directed northward.}
\label{box}
\end{figure}

\section{The coupled equations}

Having described these simple models for atmospheric and oceanic circulation, and the physical interpretation of their variables, we can now identify three mechanisms by which they interact:
\begin{enumerate}
\item The pole-equator temperature contrast is supposed to be in permanent equilibrium with the wind current $x$, i.e. we put $T_{a}\propto x$. Also, the forcing by temperature contrast in \rf{lor84x} is adjusted, so we put $F\rightarrow F_{0}+F_{1}T$. This expresses the simplest geostrophic equilibrium: a north-south temperature gradient which drives a east-west atmospheric circulation.
\item The inhomogeneous forcing by land-sea temperature contrast in \rf{lor84y} should decrease with increasing temperature difference $T$. It is assumed that in the polar region the sea water temperature is higher than the temperature over land, while in the equatorial region it is lower. A higher temperature difference $T$ thus means a lower land-sea temperature contrast. This influence is described as a fluctuation upon a fixed forcing: $G\rightarrow G_{0}+G_{1}(T_{av}-T)$.
\item The water transport through the atmosphere is taken to be linear in the energy content of the traveling wave: $\delta\rightarrow\delta_{0}+\delta_{1}(y^{2}+z^{2})$. 
\end{enumerate}

\noindent Combining \rf{lor84x}-\rf{boxeq2} with the proposed coupling terms we obtain
\begin{align}
\dot{x}= & -y^{2}-z^{2}-a x +a(F_{0}+F_{1}T)          			\label{eq1} \\
\dot{y}= & x y- b x z - y + G_{0} + G_{1}(T_{av}-T)    			\label{eq2} \\
\dot{z}= & b x y + x z - z 			   			\label{eq3} \\
\dot{T}= & k_{a}(\gamma x - T) - |f(T,S)|T - k_{w} T  			\label{eq4} \\
\dot{S}= & \delta_{0} + \delta_{1}(y^{2} + z^{2}) - |f(T,S)|S - k_{w}S  \label{eq5} 
\end{align}
with $f$ as in \rf{flow}.  
With the coupling some new constants have been introduced. They are $T_{av}$, the standard temperature contrast between the polar and the equatorial box, $\gamma$, the proportionality constant of the westerly wind current and the temperature contrast and $\delta_{1}$, a measure for the rate of water transport through the atmosphere. When exploring the dynamical behaviour of the model we take $F_{1}$ and $G_{1}$ as free parameters. As motivated in the introduction, we consider small coupling to the atmosphere model. This is the case if we take $(F_{1},G_{1}) \in [0,0.1]\times [0,0.1]$.
As remarked in the previous section, we follow Roebber \cite{roeb} in scaling the parameters. In table \rf{pars} they are listed. In this scaling, one unit of time in the model corresponds to the typical damping time scale of the planetary waves. This time scale is estimated to be five to ten days.
\begin{table}[h]
\centerline{
\begin{tabular}{|l|l|l|l|} \hline
$a$ & $1/4$ & $\delta_{0}$ & $7.8 \cdot 10^{-7}$ \\ \hline
$b$ & $4$   & $k_{w}$      & $1.8 \cdot 10^{-5}$ \\ \hline
$F_{0}$ & 8 & $k_{a}$      & $1.8 \cdot 10^{-4}$ \\ \hline
$G_{0}$ & 1 & $\xi$        & $1.1 \cdot 10^{-3}$ \\ \hline
$\gamma$& 30& $\omega$     & $1.3 \cdot 10^{-4}$ \\ \hline
$\delta_{1}$& $9.6 \cdot 10^{-8}$ & $T_{av}$     & $30$       \\ \hline
\end{tabular}}
\scaption{The constants of the coupled model. With these constants the ocean and the atmosphere model have time scales that differ by a factor of about one thousand. See \cite{roeb}.}
\label{pars}
\end{table}

The system of equations \rf{eq1}-\rf{eq5} is not conservative. Energy is being added through solar heating, and dissipated in the atmosphere as well as the ocean model. It can be shown that the model is globally stable. The proof consists of defining a trapping region, and is omitted here.

\section{Bifurcations of equilibrium points}

\begin{figure}[p]
\begin{center}
\begin{picture}(365,500)(0,0)
\put(-53,235){\epsfig{file=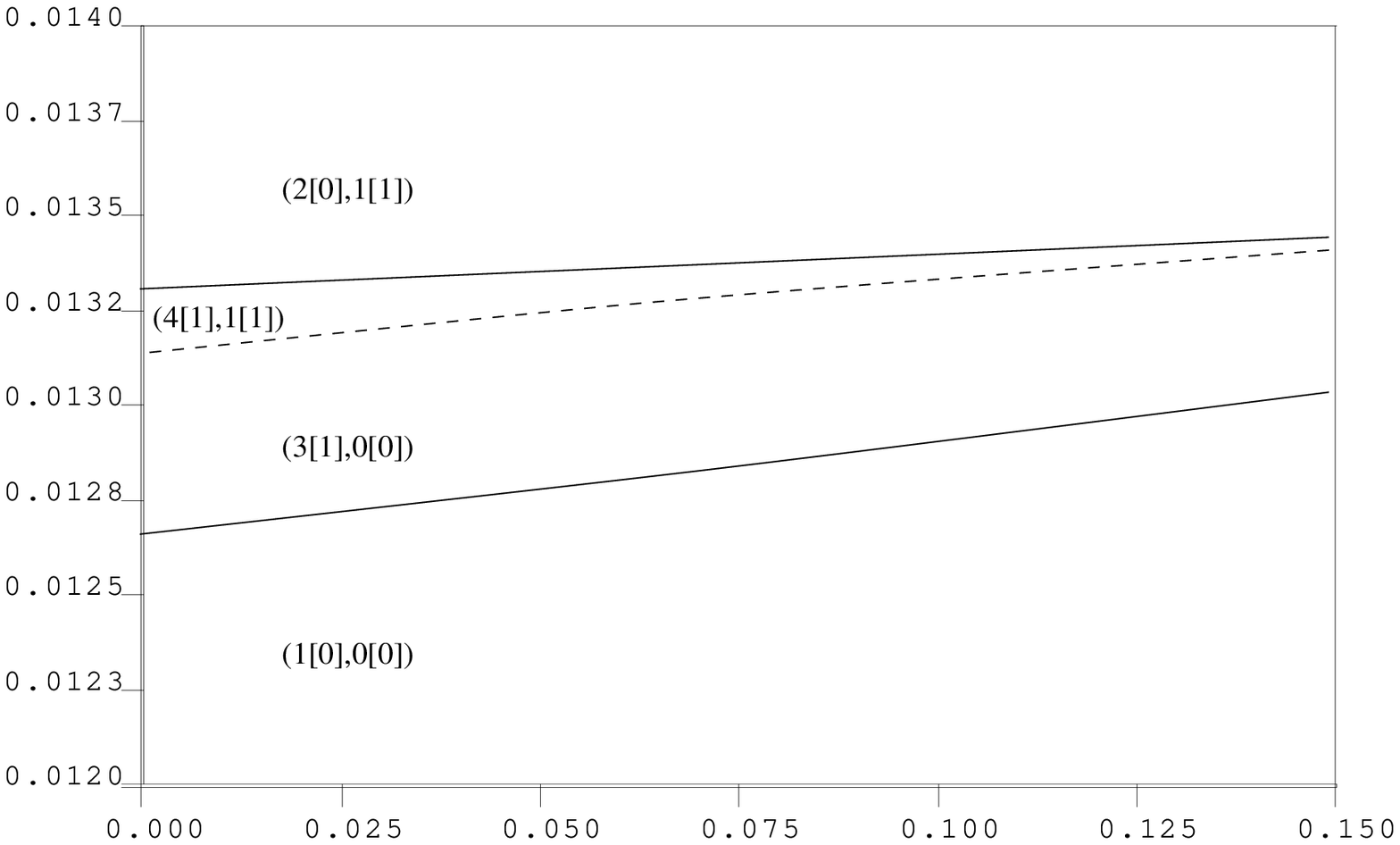,width=480pt}}
\put(-53,-37){\epsfig{file=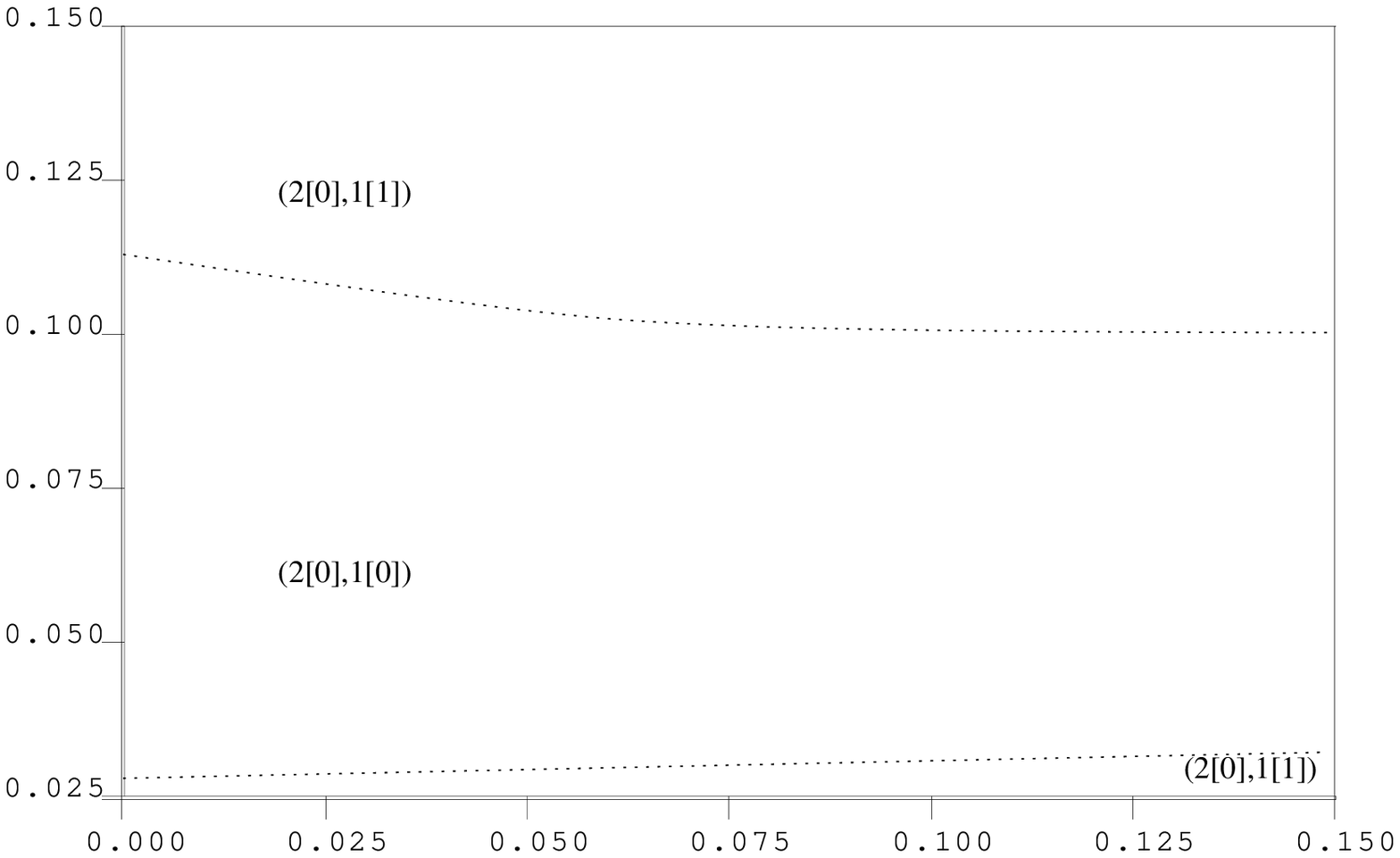,width=480pt}}
\put(337,-14){$F_{1}$}
\put(337,259){$F_{1}$}
\put(-19,210){$G_{1}$}
\put(-19,491){$G_{1}$}
\end{picture}
\end{center}
\scaption{Top: saddle node bifurcation (solid line) and the bifurcation at $f=0$, given by \rf{GFzerof} (dashed line). Bottom: Hopf bifurcation (dotted line) of the equilibrium with $f<0$. Between brackets the number of equilibria with $f>0$ and $f<0$, respectively. Between square brackets the corresponding number of {\sl stable} equilibria.}
\label{snhopf}
\end{figure}
In order to find the equilibrium points of the model, we must equate the time derivatives \rf{eq1}-\rf{eq5} to zero. By some algebraic manipulations the set of equations can be simplified, and a program like Mathematica can be used to calculate all equilibria for given parameter values, along with their spectra. In addition, the saddle node and Hopf bifurcations of these equilibria can be found using a continuation package like AUTO~\cite{doed}. On a plane in phase space, defined by $f=0$, the vector field is not differentiable. \newpage
\noindent There is an equilibrium point on this plane if
\begin{equation}
G_{1}= \frac{-G_{0} \pm \sqrt{a(F_{0}+F_{1}T_{0}-x_{0})(1-2x_{0}+(1+b^{2})x_{0}^{2})}}{T_{av}-T_{0}}.
\label{GFzerof}
\end{equation}
with equilibrium values $x_{0}=(\delta_{0}+a \delta_{1} F_{0})(k_{a}+k_{w})\xi /(\omega k_{w} k_{a} \gamma+ a \delta_{1}\xi [k_{a}+k_{w}-F_{1}\gamma k_{a}])$  and $T_{0}=k_{a}\gamma x_{0}/(k_{a}+k_{w})$. 
On the curve in parameter space, defined by \rf{GFzerof}, a bifurcation occurs. When crossing it, increasing $G_{1}$, two equilibrium points appear, one with a positive value of $f$, and one with a negative value. The latter is stable. In fact, for any $G_{1}$ greater than the right hand side in \rf{GFzerof} there is an attracting equilibrium or periodic solution on which $f$ is negative, i.e. the THC is inverted.

The results of the bifurcation and stability analysis are shown in figure \rf{snhopf}. The stability of the equilibrium points is indicated in the diagrams. As can be seen, only in a small window in parameter space there exist a stable equilibrium with positive flow. The attractors which arise in the regime with positive flow, i.e. a temperature driven THC, are either periodic or chaotic. The behaviour in this regime is more complex than in the regime with a salinity driven THC. A reason for asymmetry may be that the coupling through $\delta_{1}$ is rather weak. Experiments with more realistic models indicate that the water vapour transport through the atmosphere, represented by this constant, should be made a function of the temperature difference $T$, as the air temperature influences the efficiency of the transport.

\section{Chaotic attractors}

The bifurcation diagrams given so far only display bifurcations of equilibrium points. It turns out that in a large parameter range, there is an abundance of periodic orbits that undergo saddle node, torus and period doubling bifurcations. In the uncoupled Lorenz model there exists a codimension two point that acts as an organising center for the bifurcation diagram. At such a point, normal form theory can be employed to find the local bifurcation structure. By continuation techniques information can be gained about the global bifurcation structure. Such an analysis is described in \cite{shil}. The absence of such a point in the coupled model makes it quite hard, if not impossible to find and characterise the complete bifurcation structure. Instead we can do brute force integrations in order to classify the behaviour of the model. 

It is found that for many parameter values the behaviour is chaotic. Using the algorithm described by Wolf {\sl et al} \cite{wolf} we can approximate the Kaplan-Yorke dimension of the chaotic attractors. For several parameter values it is found to be about $4.3$, compared to the typical Kaplan-Yorke dimension of about $2.4$ for the Lorenz-84 model. This quantity, however, only characterises the geometry of the attractor. Even if there is very little variability in the degrees of freedom we add from the ocean model, the Kaplan-Yorke dimension is increased by nearly one for each degree of freedom. A way to keep track of the dynamics on the attractor, is to calculate the correlation dimension. If there is little variability in the degrees of freedom we add, the attractor of the combined system will be dynamically 'flat' and the correlation dimension will not increase as much as the Kaplan-Yorke dimension. Indeed, for the uncoupled Lorenz model the correlation dimension is typically about $2.3$ \cite{anas}, compared to $3.4 \pm 0.2$ for the coupled model. In other words, the attractor of the coupled system is much more inhomogeneous than that of the Lorenz system.

A chaotic attractor can coexist with an attracting equilibrium point, or a stable periodic solution, with negative flow. It does not seem to be possible for the system to change from the regime with positive flow to the regime with negative flow or vice versa repetitively. Depending on the initial conditions, one of the regimes is soon entered and stayed in for ever. Although, as mentioned above, a modification of the coupling terms might alter the behaviour in the regime with negative flow, we suspect that no mechanisms that could force such a transition are represented in the model. 

The projection of a chaotic attractor of the coupled model onto the subspace of the fast variables looks much like the chaotic attractor of the Lorenz model shown in figure \rf{lorat}. The parameter ranges in which the behaviour is chaotic are bordered by periodic regimes. The transition is through intermittency. The latter is of interest, as the intermittent behaviour seems to be enhanced by the presence of a slow time scale. We will describe in some detail how this behaviour is brought about.

\begin{figure}[p]
\begin{center}
\begin{picture}(365,500)(0,0)
\put(0,285){\epsfig{file=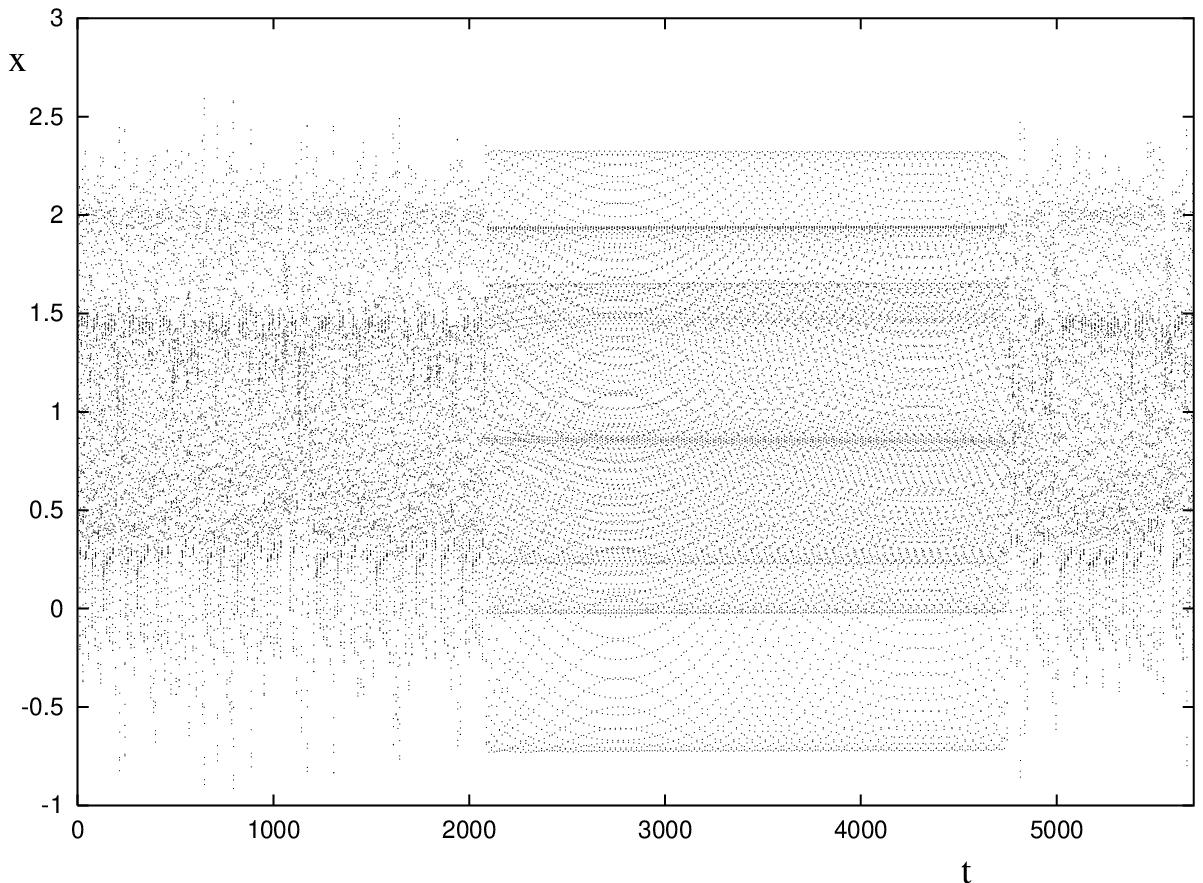}}
\put(0,5){\epsfig{file=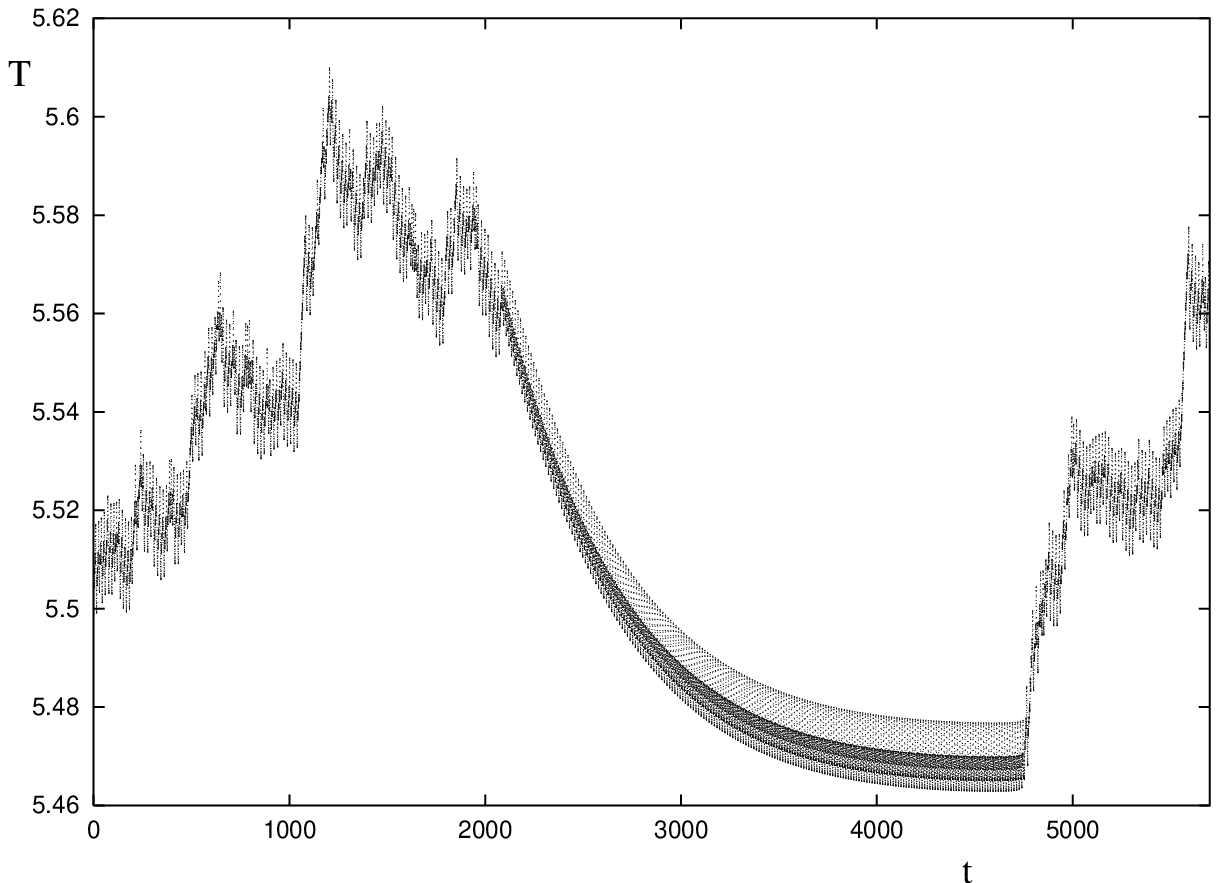}}
\end{picture}
\end{center}
\scaption{Time series at parameter values $(F_{1},G_{1})=(0.021685,0.01)$. Top: $x$, bottom: $T$. The apparent curves are due to aliasing.}
\label{interm1}
\end{figure}
\begin{figure}[p]
\begin{center}
\begin{picture}(365,500)(0,0)
\put(-33,285){\epsfig{file=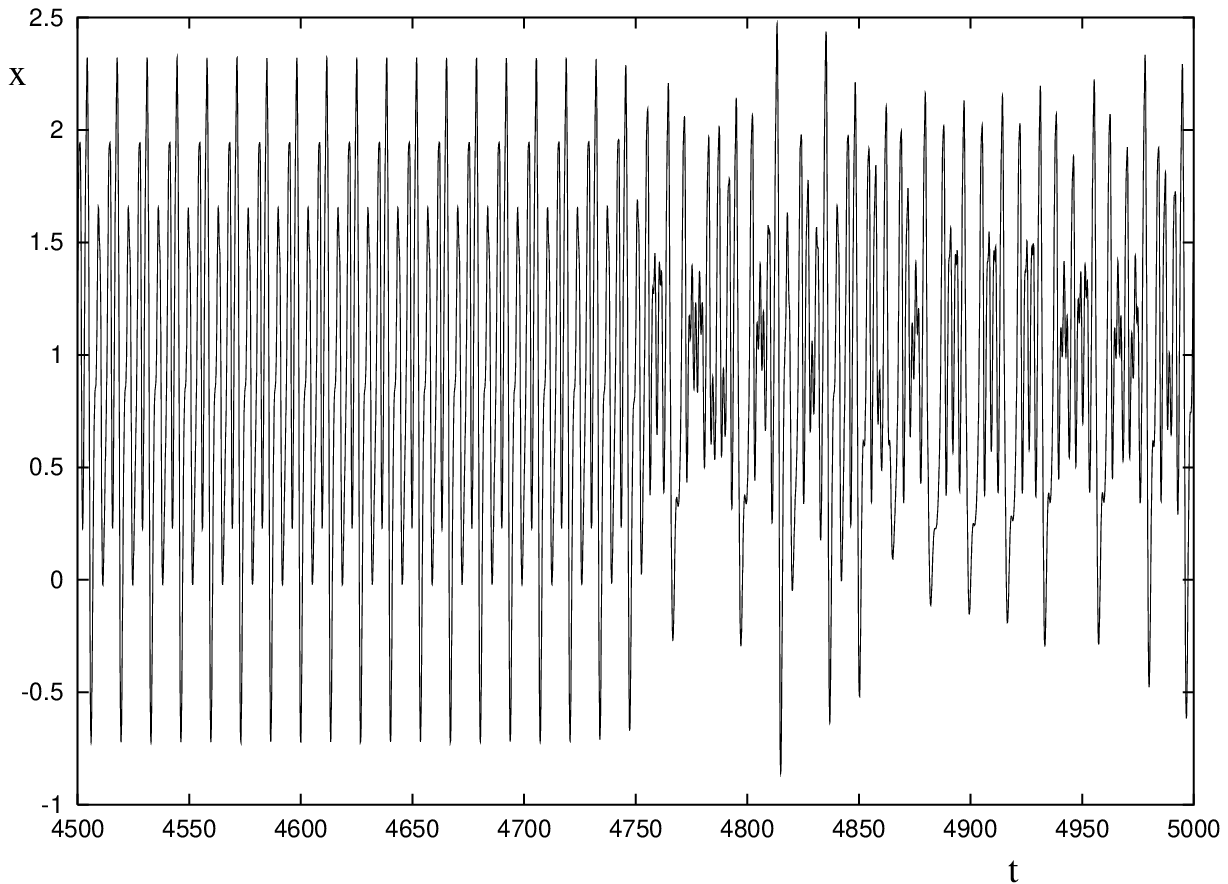}}
\put(-33,4){\epsfig{file=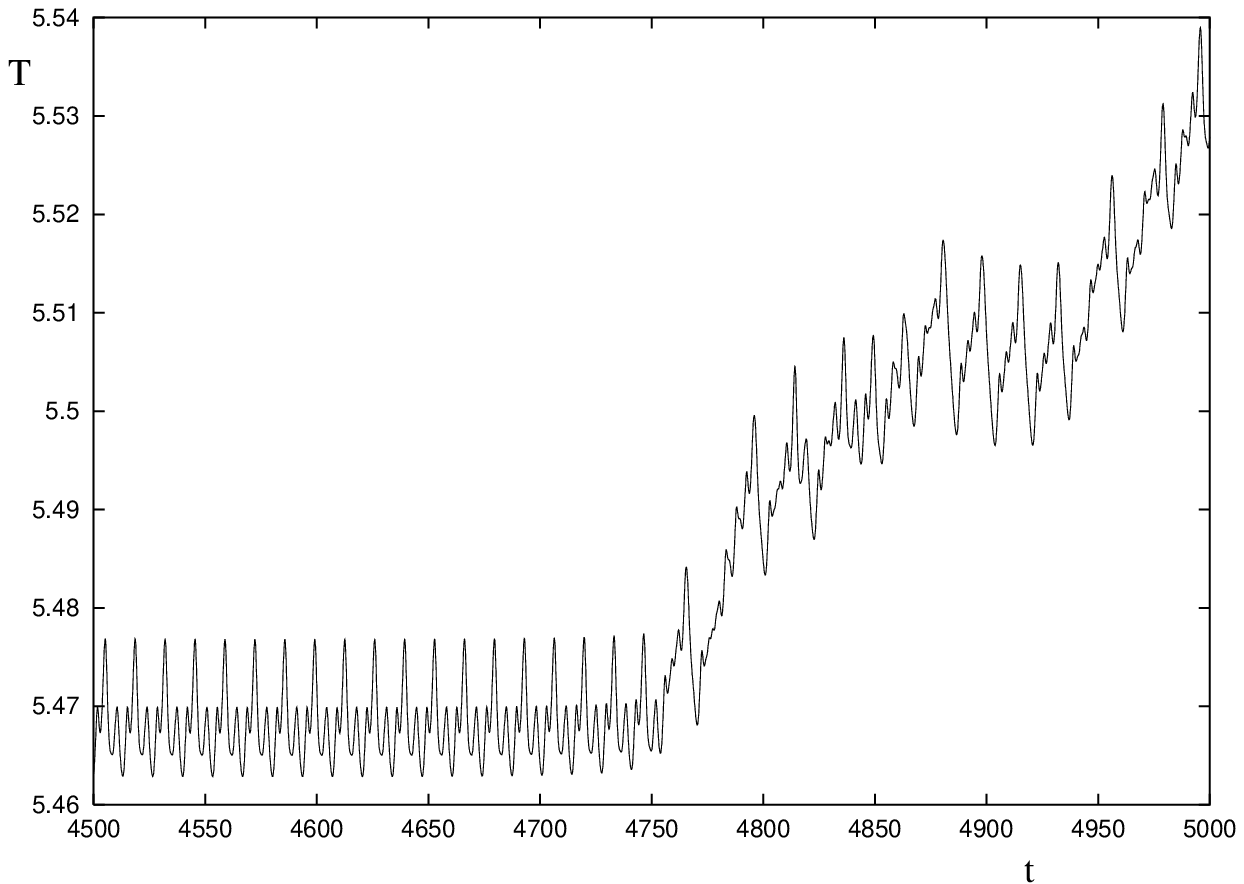}}
\end{picture}
\end{center}
\scaption{The transition to chaotic motion, enlarged from figure \rf{interm1}.}
\label{interm2}
\end{figure}

\section{Intermittency}

In figure \rf{interm1} a time series is shown, obtained by simply integrating the model with arbitrary initial conditions and parameter values given by $(F_{1},G_{1})=(0.021685,0.01)$. If the integration is continued, periodic intervals keep appearing. The length of those intervals is randomly distributed, but on average the (near-)periodic behaviour lasts about as long as the chaotic behaviour. This is not a transient effect, in the sense that if we integrate long enough the system will settle on periodic, weakly attracting set. It has all the properties of intermittent behaviour and is found in a range of parameter values around those used in this integration. 

There seems to be a periodic solution, which is approached during the periodic interval. Looking more closely at the transition to chaotic motion (see figure \rf{interm2}) there seem to be several periodic solutions that attract the orbit briefly. In order to identify these periodic solutions and calculate their spectra we can define a Poincar\'e map, e.g. in the section $\mathcal{S}_{x}=\{(x,y,z,T,S)\in \mathbb{R}^{5}|x=1\}$, and numerically look for its fixed points using the method of Newton-Raphson. An initial guess can be taken from integrations such as the one shown in figure \rf{interm1}. Proceeding like this we find several fixed points of saddle type. 

These periodic solutions can be continued in one of the parameters, for instance using the algorithm described in \cite{simo}. In figure \rf{perbif} the results are shown of a continuation in parameter $F_{1}$ of the periodic solution approached in the periodic regime in figure~\rf{interm1}. There is quite a large number of saddle node bifurcations in this continuation as well as  Neimark-Sacker bifurcations and period doublings, which are not shown in the picture. Doing the continuation for other periodic solutions, found by the method described above, we find qualitatively the same behaviour.

Looking at the spectra of the periodic solutions we find that two of the Floquet multipliers are close to, but smaller than, unity. The associated eigenvectors lie almost entirely in the subspace of the slow variables. This turns out to be a generic feature of the periodic solutions of the coupled model.

\subsection{Skeleton dynamics}

Following the Floquet multipliers of the Poincar\'e map closely near the leftmost saddle node bifurcation in figure \rf{perbif}, near which the intermittency takes place, it turns out that both branches are initially  unstable. Very close to the saddle node bifurcation a Neimark-Sacker bifurcation occurs, at which two multipliers cross the unit circle as a complex pair. 
Past the Neimark-Sacker point, the branch with the higher period is stable and periodic behaviour sets in. To the left of this point the behaviour is intermittent, and remains so left of the saddle node bifurcation. This is because the saddle node bifurcation is local. Some distance away from the bifurcating orbit in phase space, the vector field remains essentially the same. Thus the `ghost' of the periodic solutions still influences the global dynamics. This effect was labeled `skeleton dynamics' by Nishiura in a recent preprint on transient phenomena in partial differential equations \cite{nish}. 

The farther away from the saddle node point the parameters are chosen, the less the influence of the skeleton structure. This notion can be quantified by measuring the length of the periodic intervals, or rather its distribution, for a number of parameter values. To obtain these data, integrations of $5\times 10^6$ in units of $t$ (about $6.8\times 10^4$ years) were done, during which more that $600$ periodic intervals were registered. This was done by tracing approximate recurrences of points under the Poincar\'e map on $\mathcal{S}_{x}$. In the chaotic as well as in the periodic phase $x$ fluctuates about unity, so that the Poincar\'e map is always defined. In contrast, the mean value of $T$ and $S$ differs significantly between the two phases.

From the distributions at different parameter values we can calculate the expectation value, denoted by $l$, and plot it against the parameter. Thus, we can see the decreasing effect of the skeleton structure as we go farther from the saddle node point. Another way to see this is to look at the relative amount of time spent in the periodic regime, denoted by $\tau_{per}$. Both these measures are plotted in figure \rf{skel}.

\subsection{The theory of intermittency}

\begin{figure}[p]
\begin{center}
\begin{picture}(365,500)(0,0)
\put(-34,240){\epsfig{file=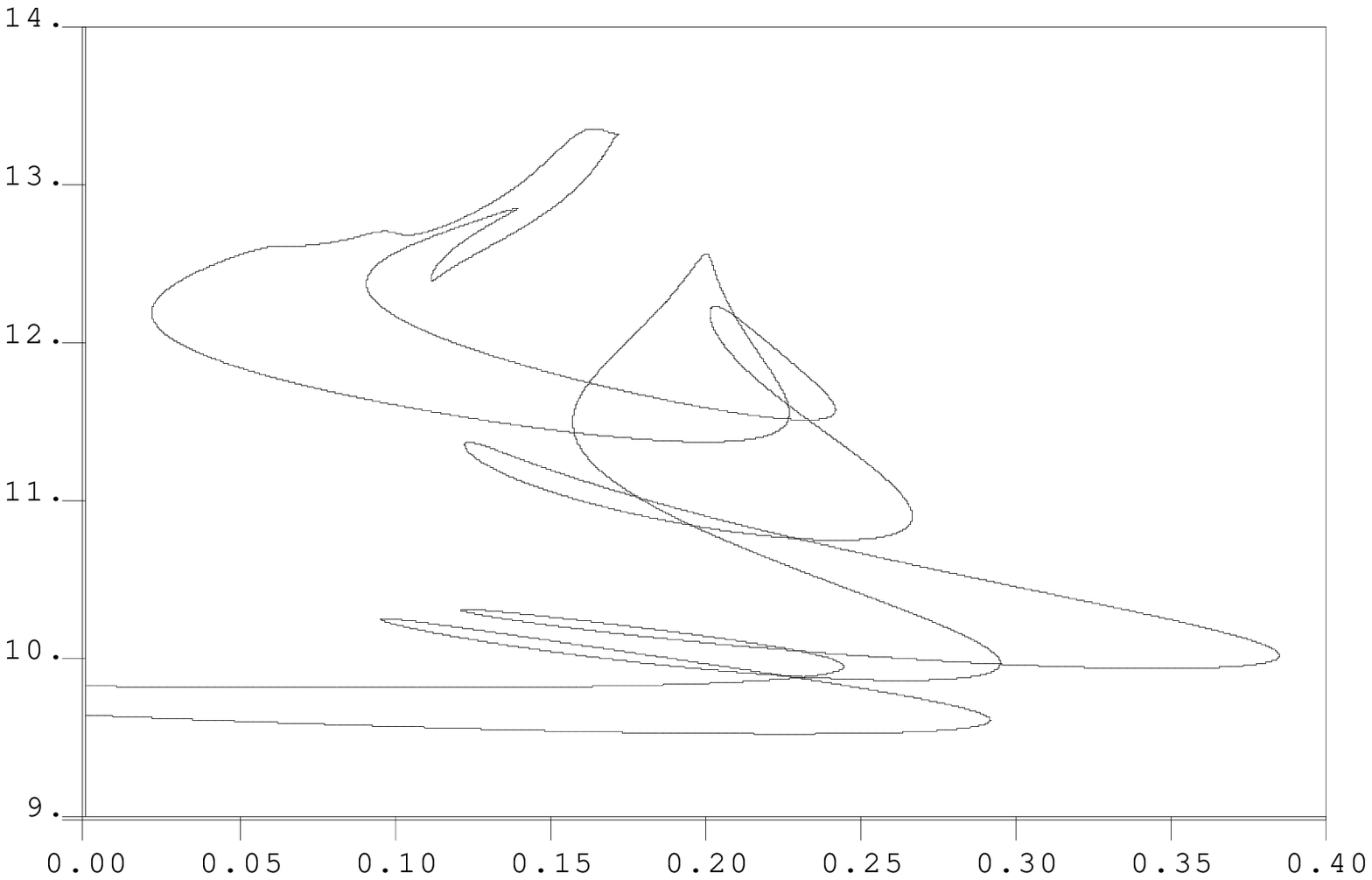,width=420pt}}
\put(-25,-19){\epsfig{file=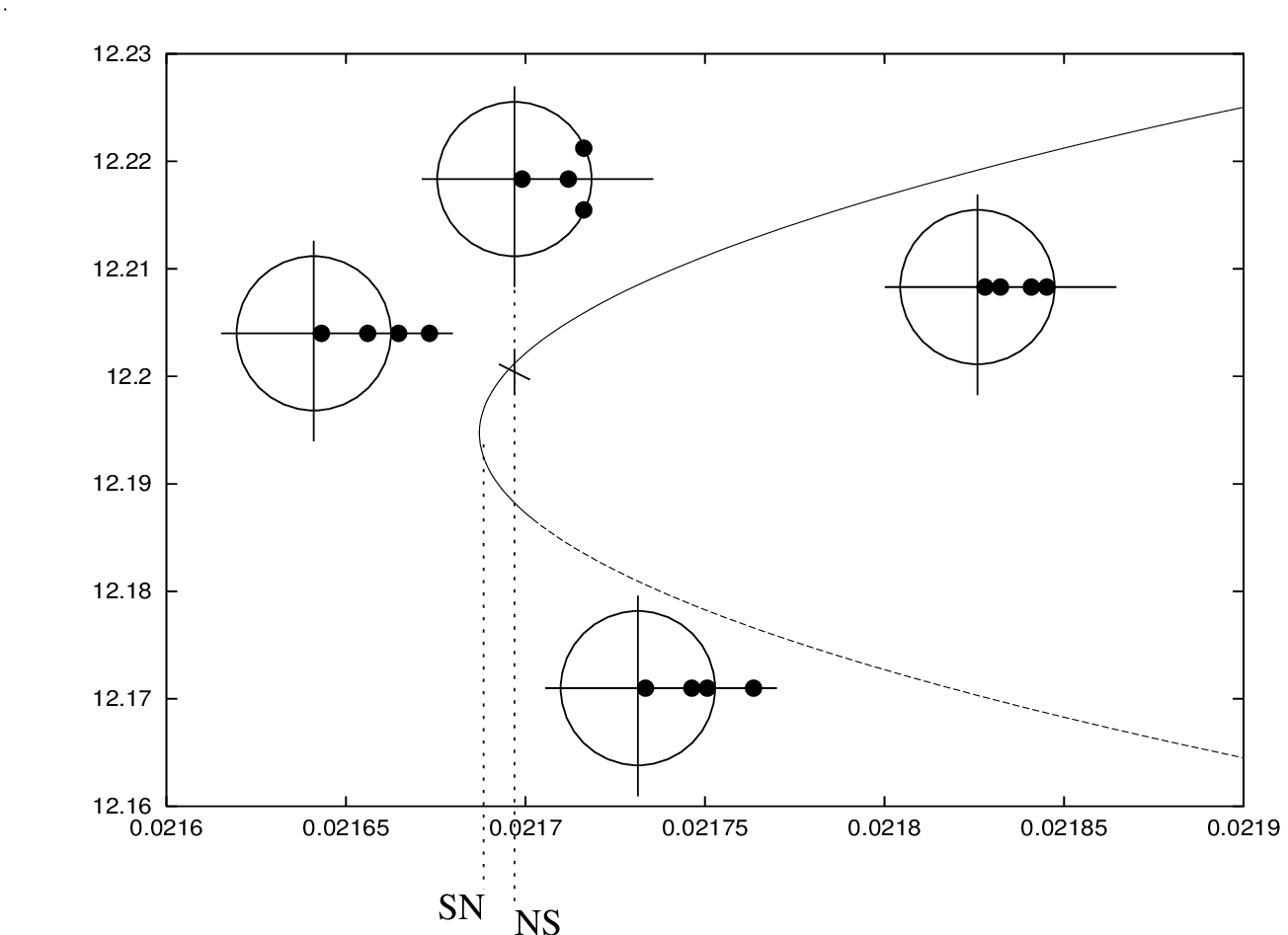,width=374pt}}
\put(305,-4){$F_{1}$}
\put(305,254){$F_{1}$}
\put(0,243){Period}
\put(0,490){Period}
\end{picture}
\end{center}
\scaption{Continuation of the periodic solution approached during the periodic regime in \rf{interm1}. Top: AUTO results, torus bifurcations and period doublings not shown. Bottom: enlargement of the leftmost saddle node bifurcation, at which the intermittency takes place. The Floquet multipliers are drawn in the complex plane. The bifurcation points are indicated by SN for saddle node and NS for Neimark-Sacker. This picture was obtained applying the algorithm described in \cite{simo}. Not shown is the period doubling at which the upper branch becomes unstable again (see figure \rf{2dcont}).}
\label{perbif}
\end{figure}
\begin{figure}[p]
\begin{picture}(365,500)(0,0)
\put(0,269){\epsfig{file=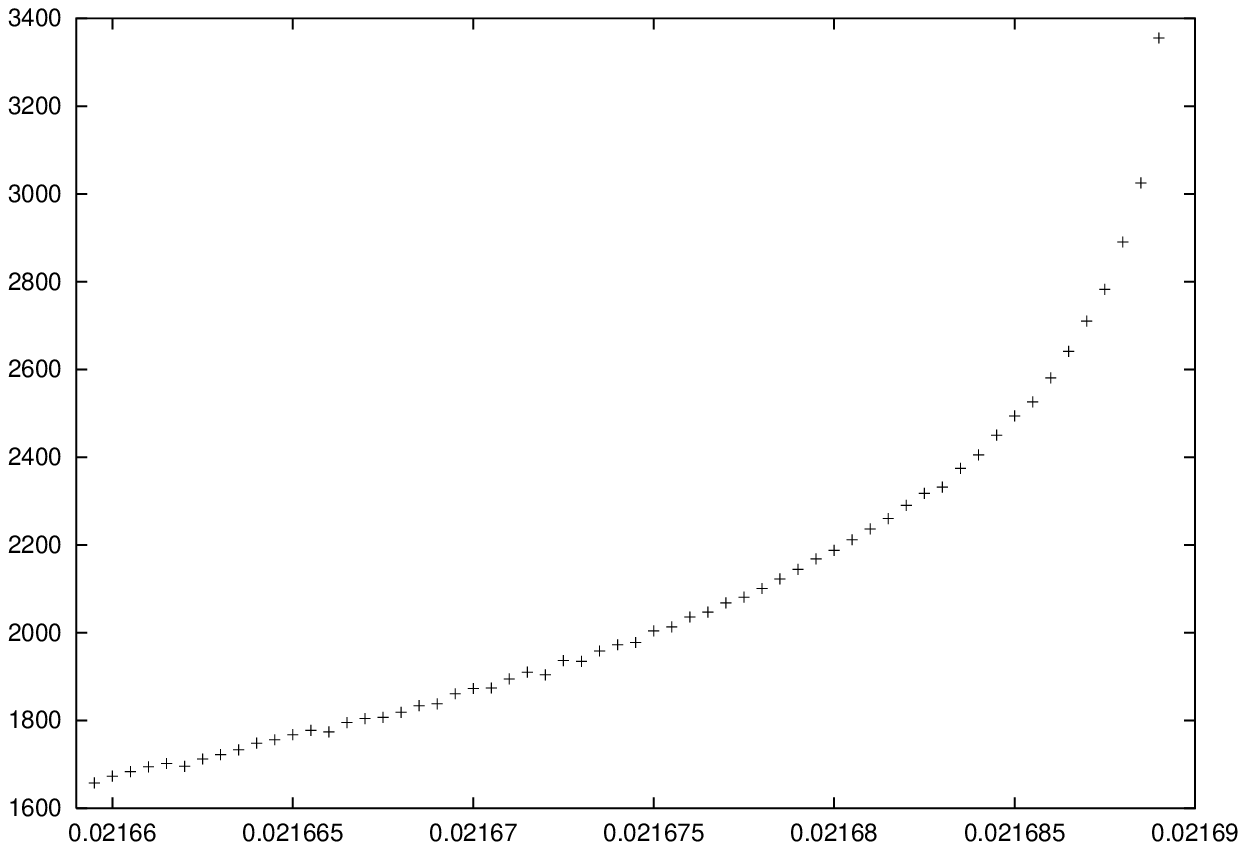}}
\put(0,2){\epsfig{file=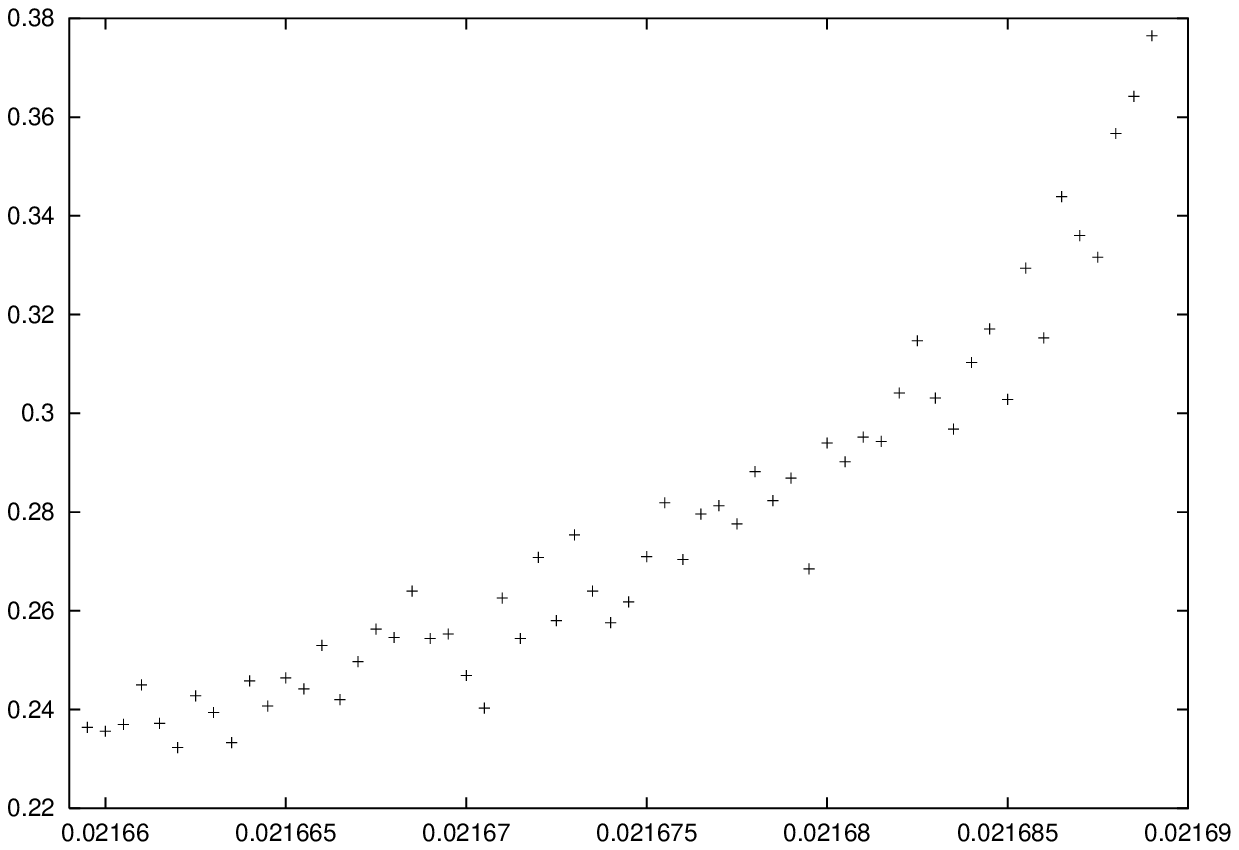}}
\put(325,260){$F_{1}$}
\put(325,-4){$F_{1}$}
\put(2,490){$l$}
\put(0,225){$\tau_{per}$}
\end{picture}   
\scaption{Top: the expectation value $l$ of the length of the periodic intervals as a function of $F_{1}$ for $G_{1}=0.01$. Bottom: $\tau_{per}$, the relative amount of time spent in the periodic regime, during a $2.8\times 10^4$ years integration run.}
\label{skel}
\end{figure}

The idea of studying the distribution of the length of the periodic intervals was first phrased by Pomeau and Manneville \cite{pome}. They made theoretical predictions of the dependence of the expectation value on the bifurcation parameter near its critical value. Three types of intermittency are distinguished, one for each generic bifurcation of a periodic solution. They are named type I, II and III for saddle node, Neimark-Sacker and period doubling bifurcations, respectively. Therefore, the intermittent behaviour described here is labeled type II. The theoretical prediction for the order of the divergence is $l\propto \ln 1/\epsilon$, where $l$ is the expectation value of the length of a periodic interval, and $\epsilon=F_{NS}-F_{1}$ the distance from the bifurcation point. In \cite{pome} it is remarked that numerical experiments suggest a power law scaling as $l\propto \epsilon^{-\alpha}$, where $\alpha\approx 0.04$. Our experiments yield a power law scaling with $\alpha\approx 0.06$, in reasonable agreement.

\subsection{Chaotic bursts}
\begin{figure}[p]
\begin{picture}(365,500)(0,0)
\put(0,5){\epsfig{file=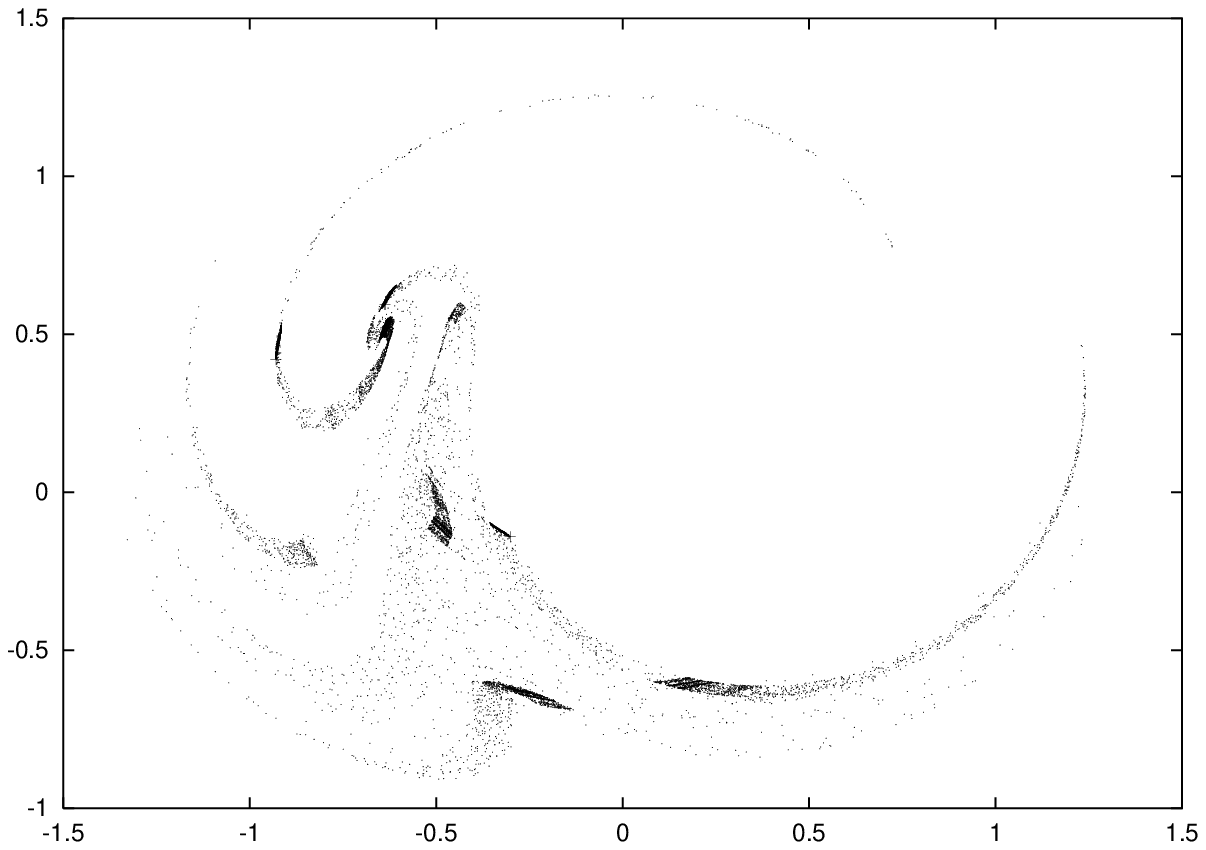,width=361pt}}
\put(-20,274){\epsfig{file=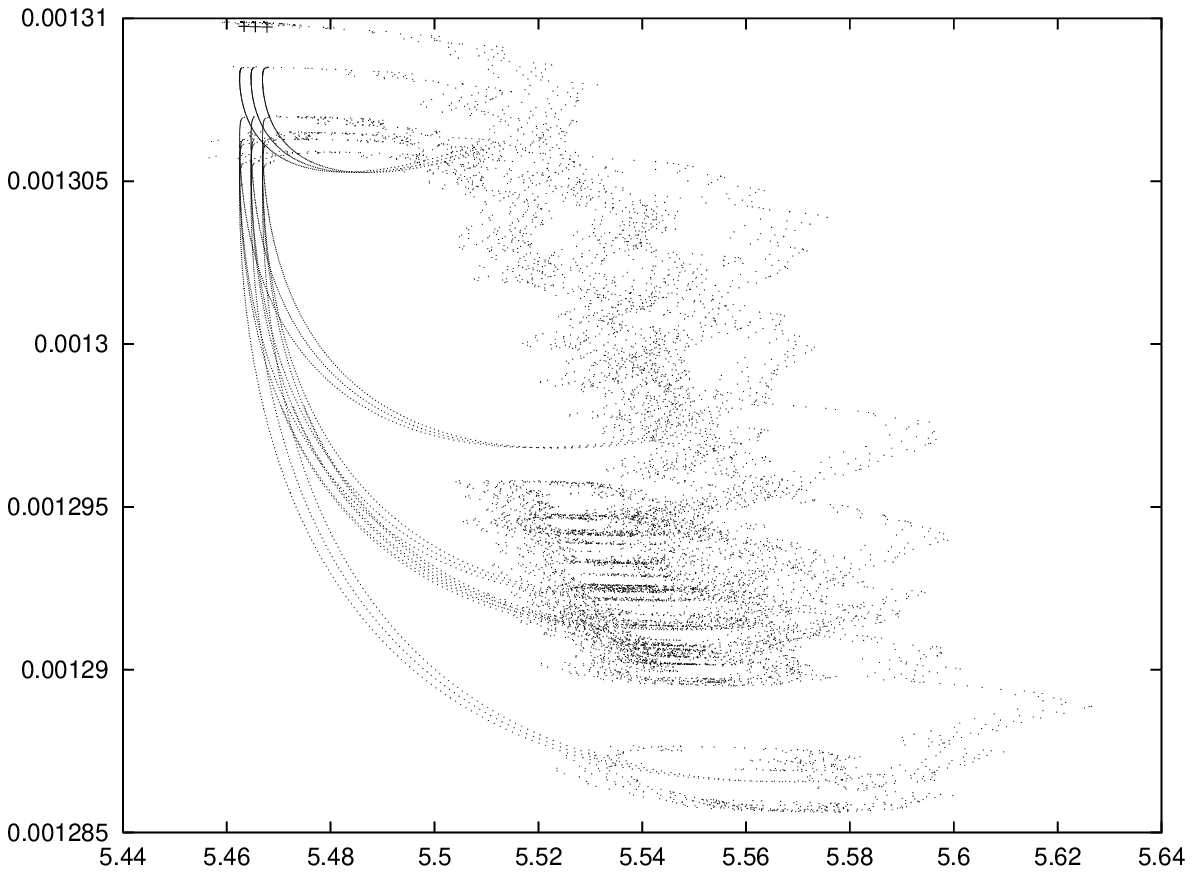,width=380pt}}
\put(15,504){$S$}
\put(325,259){$T$}
\put(325,-3){$y$}
\put(15,216){$z$}
\end{picture}   
\scaption{Poincar\'e section $\mathcal{S}_{x}$, at $(F_{1},G_{1})=(0.021685,0.01)$. Top: projection onto the $(T,S)$-plane, bottom: projection onto the $(y,z)$-plane. The intersections of the unstable periodic solution have been marked with a cross.}
\label{xcon}
\end{figure}

The missing ingredient in this description of the intermittent behaviour is the understanding of the chaotic bursts. During these bursts the motion cannot be distinguished from fully developed chaos, at least not by the eye. Numerically calculated Lyapunov exponents converge to the same values as found at fully developed chaos, two being positive. The conjecture is, that during this seemingly chaotic motion, the orbit is trapped by the numerous periodic solutions and their stable and unstable manifolds, that may intersect in a complicated way. However, the stable manifold of one of the periodic orbits that stem from the saddle node bifurcation described above, or, left of the bifurcation point, its ghost, partly lies in this tangle. The phase point moves around more or less randomly (locally, nearby orbits diverge) until it get trapped by this stable manifold and is attracted toward the weakly unstable periodic solution or its ghost. This is also referred to as the reinjection process. 

If we continue the Neimark-Sacker bifurcation in two parameters, we see that the reinjection process breaks down outside a certain range. Decreasing $G_{1}$, another periodic solution becomes stable and the behaviour becomes periodic. Increasing $G_{1}$ beyond some threshold, we see a one way transition to chaos, such as described in \cite{kara}. In figure \rf{2dcont} the results of a continuation in two parameters is shown, indicating the range in which intermittent behaviour can be observed. There are several such windows in parameter space.

Looking at the spectra near the saddle node in figure \rf{perbif}, one can see that the upper branch has a stable manifold of dimension two and the lower branch of dimension three. Both have one rapidly contracting dimension (multiplier close to zero), whereas the other directions are slowly contracting. As mentioned before, the periodic solutions of this system typically have two real eigenvalues close to unity. They are related to the long time scales of the ocean model variables. Thus, the slow dynamics of the ocean model enhances the length of the periodic intervals. Other periodic solutions also attract the orbit briefly, that is why they can be found in the first place. But they have at least one multiplier in the order of ten to one thousand, so that the time the orbit can be expected to linger in their neighbourhood is of the same order as their period or smaller. 

A nice illustration of the above conjecture is given by pictures of Poincar\'e sections. As these sections are four dimensional, we can only look at some projections. Shown here are the section $\mathcal{S}_{x}$ as defined above, projected onto the $(y,z)$-plane and the $(T,S)$-plane (figure \rf{xcon}). The periodic solution itself is a fixed point of the third iterate of the Poincar\'e map. In the pictures its intersections have been marked by a cross.

The integration from which the section $\mathcal{S}_{x}$ was obtained was started near the periodic solution, with a small perturbation in the unstable direction, so as to get an indication of the shape of the unstable manifold. In figure \rf{xcon}(bottom) it is clearly visible how the orbit comes to the chaotic region and wanders around until it gets trapped by the stable manifold again and slowly approaches the periodic solution. The integration was stopped after a few approaches in order to get a clear picture. The same data are plotted in a other projection in figure \rf{xcon}(top). Here, the typical shape of the Lorenz-attractor is clearly visible. The intersections of the periodic solution form tiny near-periodic islands in a chaotic sea.
\begin{figure}[h]
\begin{picture}(375,250)(0,0)
\put(-20,-25){\epsfig{file=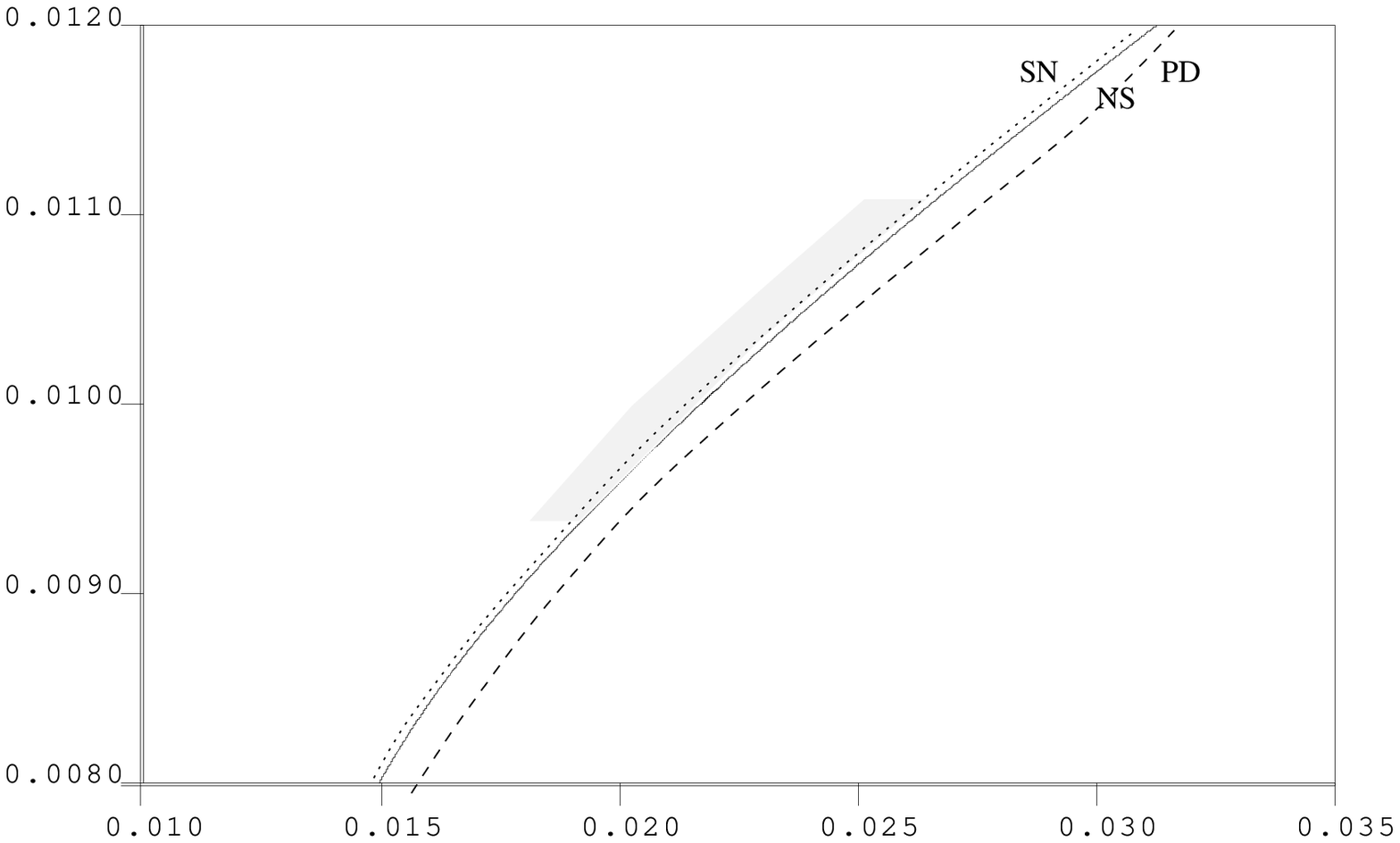,width=450pt}}
\put(335,0){$F_{1}$}
\put(20,205){$G_{1}$}
\end{picture}
\scaption{Continuation in two parameters of the saddle node (dots), the Neimark-Sacker point (solid) and the period doubling (dashed). Right of the period doubling the behaviour is chaotic, between the period doubling and the Neimark-Sacker line it is periodic, and left of the Neimark-Sacker line it is intermittent in the shaded region, chaotic above and left of the shaded region and periodic below it.}
\label{2dcont}
\end{figure}

\section{Conclusion}

When varying the coupling parameters of the model, we find equilibrium points as well as periodic solutions and chaotic attractors. The presence of competing attractors with a different orientation of the THC is inherited from the ocean box model. The chaotic attractor of the coupled model is rather inhomogeneous. In the chaotic regime, the atmospheric dynamics is dominant and there is little variability in the oceanic variables. 

The transition from periodic to chaotic motion can be intermittent. The intermittent behaviour is found near a Neimark-Sacker bifurcation, in which a periodic solution loses its stability. It persists beyond the point where the unstable periodic solution disappears in a saddle node bifurcation. Following the approach of Pomeau and Manneville \cite{pome}, the average length of a periodic interval has been measured as a function of the bifurcation parameter. Its divergence as the bifurcation parameter approaches the Neimark-Sacker point obeys a power law scaling, in agreement with the results of Pomeau and Manneville.

There are numerous periodic solutions in phase space, which share the property that two of their Floquet multipliers are smaller than, but close to, unity. These are related to the slow evolution of the oceanic variables. The time scale of the periodic intervals during the intermittent behaviour is set by this slow evolution. Thus, the intermittent behaviour is enhanced by the coupling to the slow ocean model. 

The bifurcation scenario leading to intermittency is found in several places in parameter space. It involves only generic, codimension one, phenomena. Therefore it might be expected in other chaotic slow-fast systems. Whether or not it plays a role in more realistic climate models, of higher dimension, remains to be investigated.

\section{Acknowledgments}

Part of the work by the first author was done at the faculty of applied and analytical mathematics of the University of Barcelona. The hospitality and helpfulness of Carles Sim\'o and his group is gratefully acknowledged. Contributions were also made by F.A. Bakker and G. Zondervan at the KNMI. This work is part of the NWO project `a conceptual approach to climate variability' (nr. 61-620-367).

\bibliography{pre}

\begin{thebibliography}{10}
\bibitem{not1}
This is a plane on which the variation of the Coriolis force with lattitude is
  taken into account in first order. It is often used when describing
  midlatitude atmospheric dynamics.
\end{thebibliography}
\bibliographystyle{siam}

\end{document}